\documentclass[runningheads]{llncs}
\usepackage[T1]{fontenc}
\usepackage{graphicx}
\usepackage{subcaption}
\usepackage[square,numbers,sort&compress]{natbib}
\setcitestyle{square, comma, numbers, sort&compress}

\begin{document}

\title{SensPS: Sensing Personal Space Comfortable Distance between Human-Human Using Multimodal Sensors}
\titlerunning{SensPS: Sensing Personal Space Using Multimodal Sensors}

\author{
Ko Watanabe~\inst{1}\orcidID{0000-0003-0252-1785} \and \\
Nico Förster~\inst{2}\orcidID{0009-0005-6312-3096} \and \\
Shoya Ishimaru~\inst{3}\orcidID{0000-0002-5374-1510}
}
\authorrunning{Watanabe et al.}

\institute{
German Research Center for Artifcial Intelligence(DFKI), Kaiserslautern, Germany \and 
RPTU Kaiserslautern-Landau, Kaiserslautern, Germany \and
Osaka Metropolitan University, Osaka, Japan
}

\maketitle

\begin{abstract}
Personal space, also known as peripersonal space, is crucial in human social interaction, influencing comfort, communication, and social stress. 
Estimating and respecting personal space is essential for enhancing human-computer interaction (HCI) and smart environments.
Personal space preferences vary due to individual traits, cultural background, and contextual factors. Advanced multimodal sensing technologies, including eye-tracking and wristband sensors, offer opportunities to develop adaptive systems that dynamically adjust to user comfort levels. 
Integrating physiological and behavioral data enables a deeper understanding of spatial interactions.
This study aims to develop a sensor-based model to estimate comfortable personal space and identify key features influencing spatial preferences.
Here we show that multimodal sensors, particularly eye-tracking and physiological wristband data, can effectively predict personal space preferences, with eye-tracking data playing a more significant role.
Our experimental study involving controlled human interactions demonstrates, that the Transformer model achieves the highest predictive accuracy (F1 score: 0.87) for estimating personal space. 
Eye-tracking features, such as gaze point and pupil diameter, emerge as the most significant predictors, while physiological signals from wristband sensors contribute marginally. 
These findings highlight the potential for AI-driven personalization of social space in adaptive environments.
Our results suggest that multimodal sensing can be leveraged to develop intelligent systems that optimize spatial arrangements in workplaces, educational institutions, and public settings. Future work should explore larger datasets, real-world applications, and additional physiological markers to enhance model robustness.
\end{abstract}

\section{Introduction}
Personal space~\cite{nandrino2017perception, coello2021interrelation, candini2021physiological}, also known as Peripersonal space, is a fundamental aspect of human social interaction, representing the comfortable distance individuals maintain from others. 
It is a highly subjective concept shaped by cultural background, personality traits, and situational contexts~\cite{sorokowska2017preferred, beaulieu2004intercultural}.
Respecting personal space is essential for fostering interpersonal comfort, reducing social stress~\cite{sehrt2024closing}, and enhancing communication~\cite{watanabe2023engauge, watanabe2024metacognition}, especially in collaborative settings.
Understanding and accommodating personal space preferences in HCI and smart environments~\cite{cook2007smart, lim2024exploring} offer exciting opportunities to improve user experiences and optimize social interactions through adaptive, intelligent systems.

This study explores the estimation of comfortable personal space using multimodal sensor data to enable automated systems to maintain appropriate distances between individuals. 
By utilizing advanced sensing technologies, such as eye-tracking glasses and wristband sensors, we aim to develop adaptive systems that dynamically respond to users' spatial preferences, enhancing comfort and productivity in workplaces, educational institutions, and public spaces.

To achieve this, we conducted an experimental study involving pairs of participants in controlled interaction scenarios. 
Data were collected on gaze patterns, physiological indicators, and subjective comfort assessments. 
This approach comprehensively explains personal space dynamics by integrating objective sensor measurements with subjective perceptions.
Our research addresses two primary questions: 

\begin{itemize}
  \item[] RQ1: Can multimodal sensors estimate comfortable personal space?
  \item[] RQ2: What is the best model for estimating comfortable personal space?
  \item[] RQ3: What are the key features for estimating comfortable personal space?
\end{itemize}

By integrating physiological data with human-centered design principles, this research contributes to the HCI field and paves the way for intelligent systems that enhance human interactions in physical and virtual environments.

\section{Related Work}
This section reviews the related work of eye-tracking technologies, wristband sensors, cognitive state estimation with multimodal sensors, and personal space estimation.

\subsection{Eye-Tracking Technologies}
Research on eye-tracking has been carried out across various fields, including HCI and psychology. 
Studies have demonstrated that eye-tracking can be utilized to estimate factors such as confidence~\cite{bhatt2024self}, personality~\cite{berkovsky2019personality}, attention~\cite{vainio2019urban}, and cognitive load~\cite{zagermann2016cognitive}. 
Additionally, eye-tracking is employed in diverse interaction tasks like gaze-based typing~\cite{cai2023speakfaster,lystbaek2022exploring}, menu navigation~\cite{kopacsi2024exploring}, and object selection~\cite{cho2024sonohaptics}.
The choice of eye-tracking device varies depending on the data type and task. 
Examples include PC-mounted eye-tracking~\cite{dembinsky2024eye,dembinsky2024gaze,bhatt2024self}, eye-tracking glasses~\cite{bykowski2018automatic,meteier2023enhancing}, head-mounted displays~\cite{minakata2019pointing}, and webcam-based eye-tracking~\cite{ankur2024appearance,shah2024webcam,shah2024eyedentify}.

Each eye-tracking methodology has its distinct strengths and weaknesses~\cite{carter2020best,shi2024bibliometric}.
For example, PC-mounted eye-tracking systems are easy to set up and operate, making them user-friendly, but they lack portability and are restricted to a fixed location. 
Eye-tracking glasses provide portability and mobility but require the user to wear a specialized device. Head-mounted displays also offer portability but can be complex to set up. 
Webcam-based eye-tracking systems are easy to set up and use, yet they are less accurate than other methods.

Considering these strengths and weaknesses, selecting the appropriate eye-tracking device and task type is important when designing an eye-tracking system.
Our study focuses on estimating comfortable personal space, so we use eye-tracking glasses. 
This choice is due to their portability, allowing participants to move freely, and higher accuracy and robustness than camera-based eye-tracking.

\subsection{Wristband Sensors}
Wristband sensors are wearable devices that can measure physiological signals such as heart rate~\cite{tanaka2024concentration}, skin conductance~\cite{menghini2019stressing}, and galvanic skin response~\cite{blanchard2014automated}.
These sensors have been widely used in research and practical applications due to their ability to provide continuous, real-time data in a non-invasive manner. 
Studies have demonstrated that wristband sensors are used for various tasks, such as estimating stress~\cite{menghini2019stressing}, concentration~\cite{tanaka2024concentration}, workload~\cite{brishtel2018assessing}, and mind wandering~\cite{brishtel2020mind}.
Additionally, wristband sensors are used for interaction tasks, such as object interaction~\cite{lee2024echowrist}, text entry~\cite{funk2014using}, and step aware voice instructions~\cite{arakawa2024prism}.

This study uses a wristband sensor to measure physiological signals such as heart rate and skin conductance.
These features are significant for estimating cognitive state, which is a key factor for estimating personal space.

\subsection{Personal Space}
Personal space, also known as peripersonal space, refers to the comfortable distance individuals maintain from others, which varies subjectively among individuals~\cite{nandrino2017perception,geers2023relationship}.

Coello et al.~\cite{coello2021interrelation} explored the concepts of Interpersonal Space (IPS).
IPS is the area individuals maintain between themselves and others during social interactions. 
When this space is encroached upon, it often leads to discomfort, prompting individuals to increase the distance to regain comfort.
Our study emphasizes PPS to investigate whether physiological signals can predict personal space preferences.

Candini et al.~\cite{candini2021physiological} examined the direct link between physiological responses and the regulation of interpersonal space. 
Their study employed an ecological experimental setup where participants' skin conductance response (SCR) was measured as a confederate approached or withdrew from them at varying distances.
The results showed a significant increase in SCR when participants were exposed to close interpersonal distances, especially during approaching movements.
Additionally, the study found a functional relationship between individual SCR reactivity and preferred IPS, indicating that autonomic responses play a role in the perception and regulation of interpersonal boundaries.
This research highlights that autonomic arousal, often below conscious awareness, is vital in shaping social space preferences.

Our study uses physiological signals to estimate personal space, focusing on discomfort distance.

\section{Methodology}
In this section, we describe the device selection, data preprocessing, and machine learning and deep learning models.

\subsection{Device Selection}
\begin{figure}[t!]
  \centering
  \includegraphics[width=0.5\textwidth]{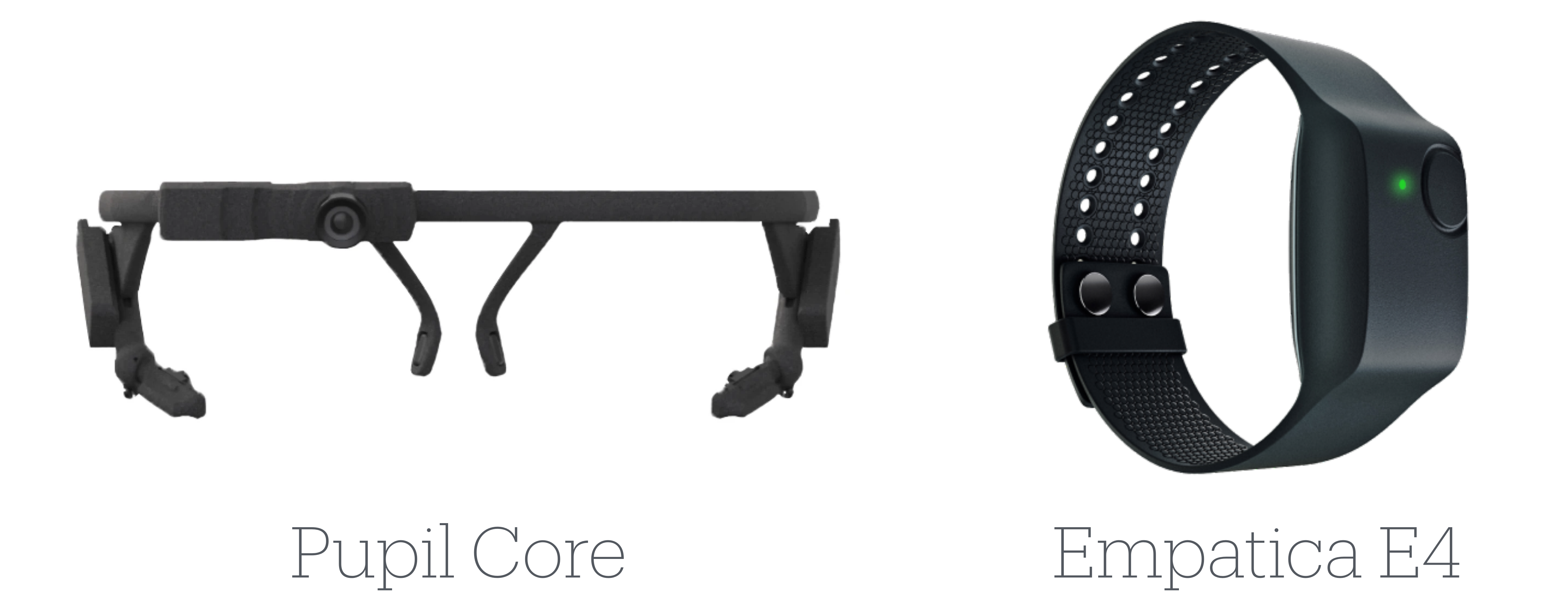}
  \caption{The devices used in this study: the Pupil Core eye-tracking glasses and the Empatica E4 wristband sensor.}
  \label{fig:device_selection}
\end{figure}

In this study, we employ the Pupil Core eye-tracking glasses~\footnote{\url{https://www.pupil-labs.com/products/pupil-core/}} and the Empatica E4 wristband sensor~\footnote{\url{https://www.empatica.com/en-int/research/e4/}}.
Figure~\ref{fig:device_selection} illustrates these devices.

The Pupil Core is a high-performance eye-tracking device renowned for its accuracy and dependability in capturing gaze data. 
It features binocular eye tracking with a sampling rate of up to 200 Hz, ensuring precise and real-time data collection. 
It is equipped with high-resolution cameras (1920x1080 pixels) and a 90-degree field of view, which provides comprehensive eye movement tracking. 
The Pupil Core supports 2D and 3D eye tracking, making it versatile for various research applications. 
Weighing only 35 grams, it is lightweight and comfortable for extended use. The adjustable headband ensures a secure fit across different head sizes. 
The device is compatible with various software tools from Pupil Labs, such as Pupil Capture and Pupil Player, facilitating data recording, visualization, and analysis. 
It offers robust connectivity options, including USB and Wi-Fi, allowing seamless integration with other devices and systems. 
This enables easy real-time data transfer and processing, enhancing research workflow efficiency. 
Overall, the Pupil Core's advanced features and user-friendly design make it an ideal choice for researchers seeking high-quality eye-tracking data in diverse settings.

The Empatica E4 wristband is an advanced wearable device for real-time physiological data collection. 
It has multiple sensors to measure various physiological signals, including heart rate, electrodermal activity (EDA), skin temperature, and motion via a 3-axis accelerometer. 
The E4 wristband includes a photoplethysmography (PPG) sensor for heart rate monitoring and an EDA sensor for skin conductance. These sensors provide high-resolution data with sampling rates of up to 4 Hz for EDA and 64 Hz for PPG. 
This ensures precise and detailed physiological measurements. The device is designed for comfort and ease of use, featuring a lightweight and ergonomic design suitable for continuous wear. 
It offers robust connectivity options, including Bluetooth, enabling seamless data transmission to connected devices. 
The E4 wristband is compatible with the Empatica Research Portal, a cloud-based platform for data management, visualization, and analysis. It allows researchers to access real-time data and download raw data for further analysis.

\subsection{Data Preprocessing}
\subsubsection{Data Cleaning:} This section details the preprocessing steps for the eye-tracking and wristband sensor data. 
First, we synchronized the data with the task duration by trimming it to the relevant period. 
For each trial, we recorded the start and end timestamps of the task. 
These timestamps were used to trim the data from the Pupil Core and Empatica E4 sensors to match the task duration. 
After trimming, we applied preprocessing to eliminate noise and artifacts. 
For the eye-tracking data, we used Pupil Labs' built-in tools for noise and artifact removal. 
For the Empatica E4 sensor data, we removed NaN values and normalized the dataset. 
We also discarded any incomplete or missing data resulting from wireless disconnections.

\subsubsection{Feature Extraction:} We extracted features from the data across statistical, temporal, and frequency domains. 
Features such as fixation, saccade, and blinks were extracted for the eye-tracking data. 
For the Empatica E4 wristband data, features included EDA, skin temperature, and accelerometer readings.

\subsubsection{Sliding Window:} A sliding window approach was employed to extract features from the data. 
The features extracted are akin to those used in Discaas~\cite{watanabe2021discaas} and Waistonbelt~\cite{nakamura2019waistonbelt}. 
The window size was set to ten seconds, with a five-second overlap. 
Features were extracted from each window segment. 
These features were then used to train machine learning and deep learning models.

\subsubsection{Labeling:} In this study, we categorized comfortable distance into two classes: comfort and discomfort. 
This was achieved by converting the Likert scale results into binary labels based on a predefined threshold.
Specifically, we classified data as comfort when the Likert score was ten and as discomfort when it was less than ten.
We aim to use the features extracted from the sliding window technique to classify these binary labels accurately.

\subsection{Machine Learning and Deep Learning Models}

\subsubsection{Machine Learning:} For machine learning, we use support vector machine (SVM), decision tree (DT), and random forest (RF) to perform binary classification. 
A support vector machine (SVM) is a supervised learning model that finds the optimal hyperplane that maximizes the margin between the two classes. 
Decision tree (DT) is a non-parametric supervised learning method for classification and regression. It splits the data into subsets based on the value of input features, creating a tree-like model of decisions. 
Random forest (RF) is an ensemble learning method that constructs multiple decision trees during training and outputs the mode of the classes for classification. It helps improve accuracy and control overfitting. 
All three methods are applied to classify the data into comfort and discomfort.

\subsubsection{Deep Learning:} VGG16~\cite{simonyan2014very} is a popular neural network for image classification, with 16 layers: 13 for convolution and 3 for fully connected tasks. It uses small filters to capture image details and pooling layers to simplify data, making it efficient and effective.

MobileNet~\cite{howard2017mobilenets} is a lightweight neural network for mobile, using fewer parameters to save resources.
It includes a base model, pooling, and a dense layer for binary classification, optimized with Adam and binary cross-entropy loss. 
MobileNetV2~\cite{sandler2018mobilenetv2} and V3~\cite{howard2019searching} are similar but more efficient, with V3 offering better performance for tasks like image classification and object detection.

The Transformer architecture~\cite{vaswani2017attention} is a deep learning model widely used for various tasks, including natural language processing. The Transformer model consists of several key components:

\begin{itemize}
    \item[] \textbf{Multi-Head Attention Layer:} This component allows the model to focus on different parts of the input sequence simultaneously. It enhances the model's ability to capture complex dependencies within the data.
    \item[] \textbf{Feed-Forward Network (FFN):} This is a simple neural network with two layers, where the first layer uses a ReLU activation function. It processes the output from the attention layer to further transform the data.
    \item[] \textbf{Layer Normalization:} This technique normalizes inputs across features, which helps stabilize and accelerate the training process by ensuring consistent input distributions.
    \item[] \textbf{Dropout Layers:} These layers are used as a regularization technique to prevent overfitting. They work by randomly setting a fraction of input units to zero during training, which helps the model generalize better to new data.
\end{itemize}

Creating a Transformer model involves constructing and compiling the model to effectively capture dependencies and relationships in the input data, making it suitable for various tasks.

\section{Data Collection}
In this section, we describe the experimental setup and data collection process.

\subsection{Participants}
We recruited ten participants for this study, aged between 21 and 28 years (Mean = 25.1). 
The sample consisted of seven males and nine females.
Participants came from a diverse range of countries, including Albania, India, Turkey, USA, China, Russia, Azerbaijan, Brazil, Germany, and Iran.
This diversity was intentional to minimize cultural bias in the results.
All participants were either undergraduate or graduate students at the University of Kaiserslautern-Landau.
Informed consent was obtained from all participants, in compliance with GDPR regulations, explaining the use of their data.

\begin{figure}[t!]
  \centering
  \includegraphics[width=0.8\textwidth]{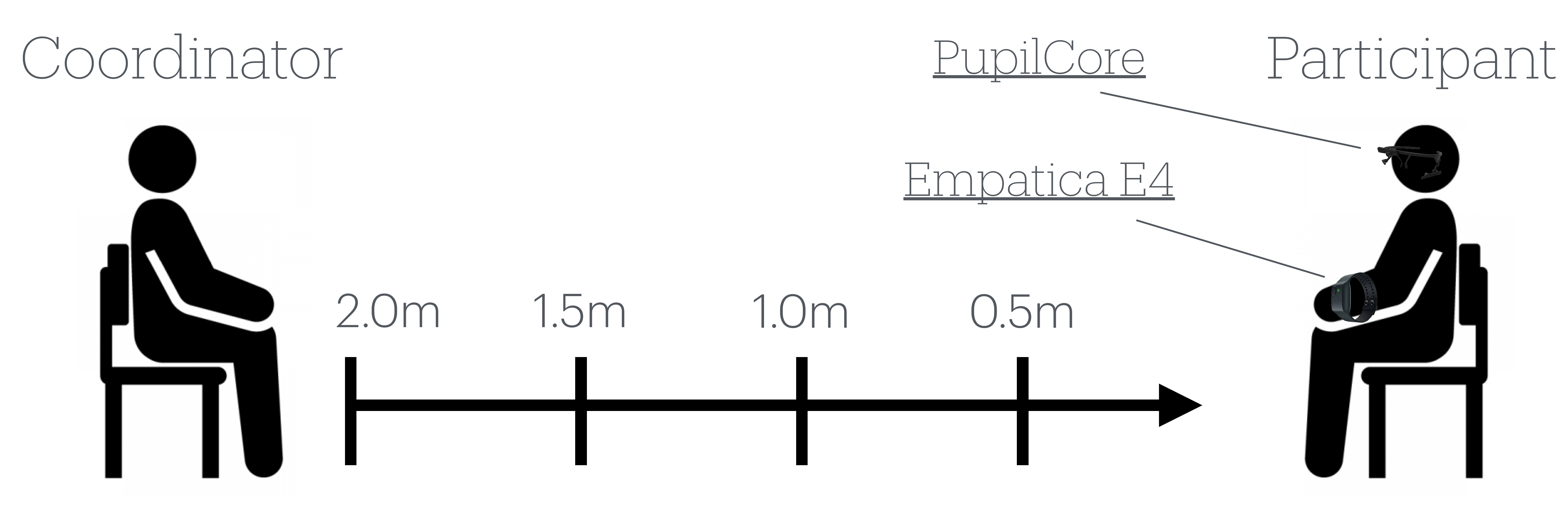}
  \caption{The experimental setup. Coordinator is the participant who is not wearing the eye-tracking glasses and wristband sensor. Participant is the participant who is wearing the eye-tracking glasses and wristband sensor. In each trial, the coordinator approaches the Participant by 0.5 meter.}
  \label{fig:experiment_setup}
\end{figure}

\subsection{Experimental Procedure}
This study employs the Pupil Core eye-tracking glasses and the Empatica E4 wristband sensor. 
Figure~\ref{fig:experiment_setup} illustrates the experimental setup. 
One participant is designated as stationary, while the other wears eye-tracking glasses and a wristband sensor. 
The Pupil Core glasses are positioned on the forehead, with cameras aligned to capture both eyes. 
The Empatica E4 wristband connects to the computer via a USB cable.

Participants first review and sign the consent form. 
They then wear the Pupil Core eye-tracking glasses and Empatica E4 wristband sensor and sit in front of a computer screen. 
A calibration and validation process follows to ensure the correct alignment of the devices. For the eye-tracking glasses, participants focus on a fixation point for alignment. 
For the wristband sensor, participants press a button to initiate data collection.

After calibration and validation, participants perform a task involving estimating comfortable personal space. 
They are seated in a chair. During each trial, the coordinator approaches the participant at distances of 2.0m, 1.5m, 1.0m, and 0.5m.
At each distance, the coordinator initiates a discussion on a topic for two minutes. 
After the discussion, the coordinator asks the participant to answer a question about the distance. 
We use a Google Form for the questionnaire~\footnote{\url{https://forms.gle/ZbgSQbaAFRpNyHVe8}}.

\section{Result and Discussion}

\begin{table}[t!]
  \centering
  \renewcommand{\arraystretch}{1.5}
  \caption{Comparison of F1 Scores across various models: Support Vector Machine (SVM), Decision Tree (DT), Random Forest (RF), MobileNet (MN), MobileNet V2 (MNV2), MobileNet V3 (MNV3), Visual Geometry Group 16 (VGG16), and Transformer.}
  \label{tab:f1_comparison}
  \begin{tabular}{l@{\hskip 0.3cm}c@{\hskip 0.3cm}c@{\hskip 0.3cm}c@{\hskip 0.3cm}c@{\hskip 0.3cm}c@{\hskip 0.3cm}c@{\hskip 0.3cm}c@{\hskip 0.3cm}c}
  \hline
  \textbf{Model} & \textbf{SVM} & \textbf{DT} & \textbf{RF} & \textbf{VGG16} & \textbf{MN} & \textbf{MNV2} & \textbf{MNV3} & \textbf{Transformer} \\ \hline
  F1 Score       & 0.53         & 0.54        & 0.67        & 0.31           & 0.63        & 0.44          & 0.33          & 0.87                 \\ \hline
  \end{tabular}
\end{table}

\begin{figure}[t!]
  \centering
  \begin{subfigure}[b]{0.48\textwidth}
    \centering
    \includegraphics[width=\textwidth]{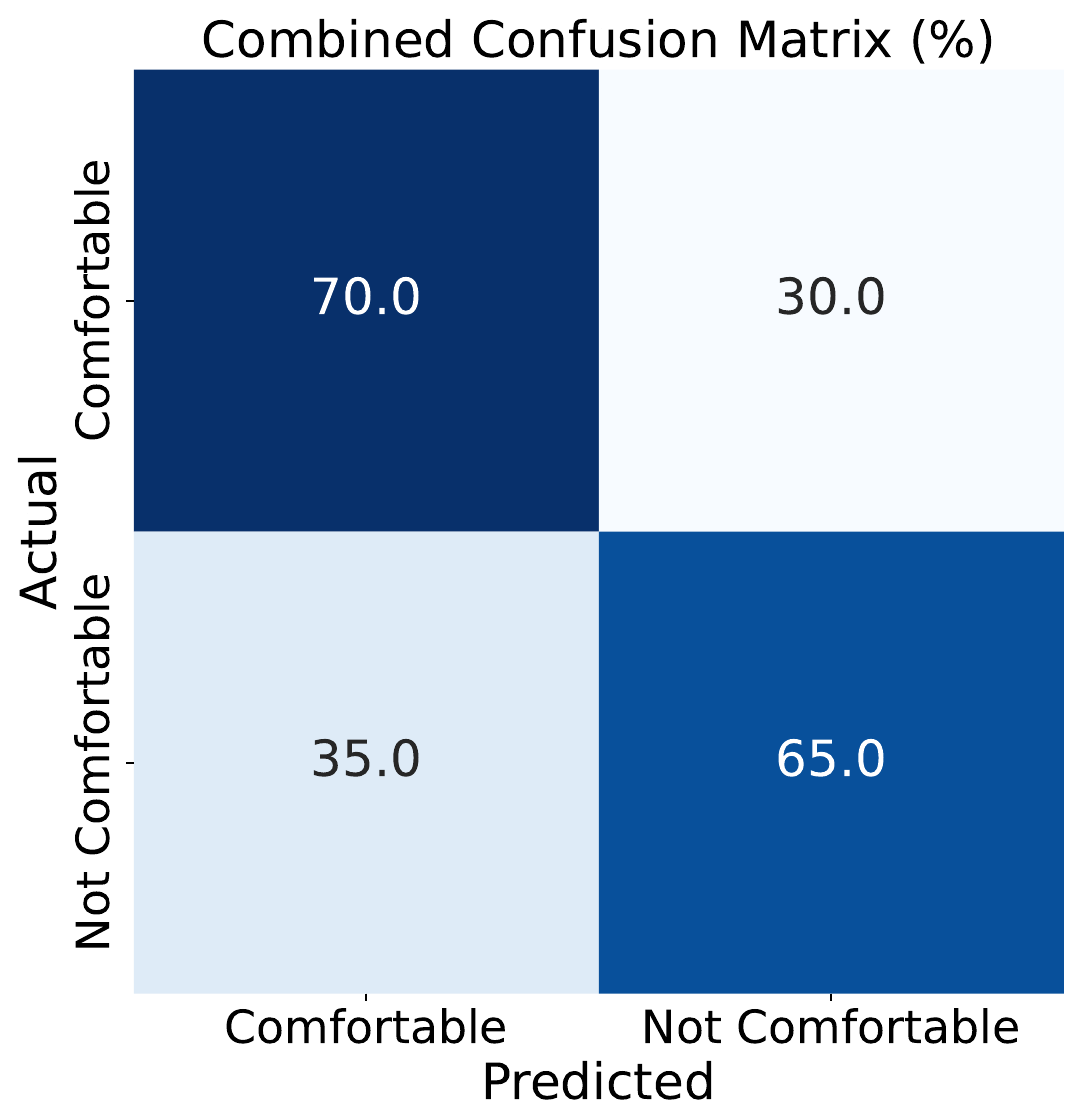}
    \caption{Confusion Matrix of Random Forest}
    \label{fig:cm_1}
  \end{subfigure}
  \hfill
  \begin{subfigure}[b]{0.48\textwidth}
    \centering
    \includegraphics[width=\textwidth]{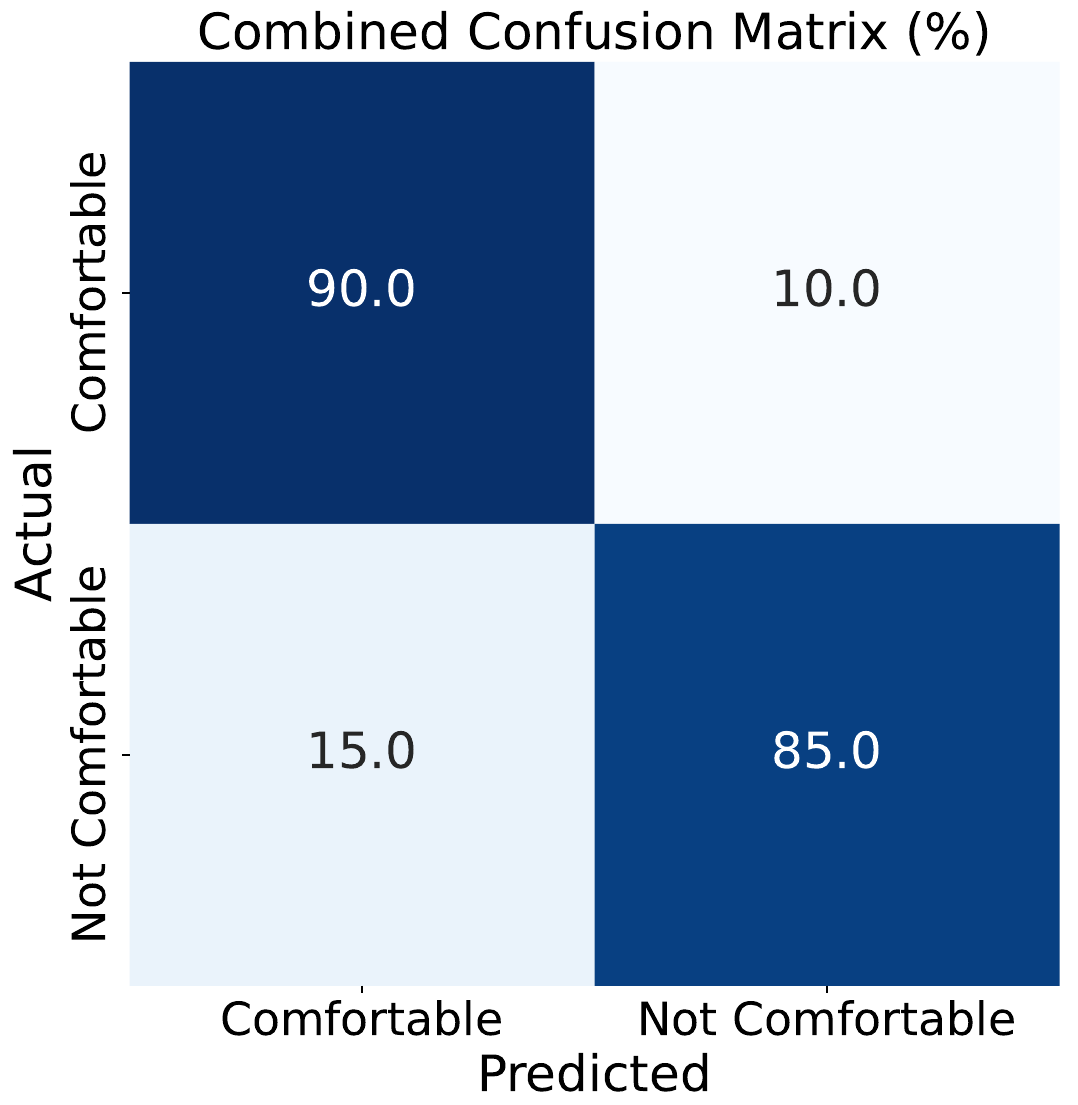}
    \caption{Confusion Matrix of Transformer}
    \label{fig:cm_2}
  \end{subfigure}
  \caption{Confusion matrices for the top-performing models: Random Forest and Transformer. The Transformer model shows a higher F1 Score in classifying instances correctly, as evidenced by its confusion matrix.}
  \label{fig:confusion_matrices}
\end{figure}

Table~\ref{tab:f1_comparison} indicates that the Transformer model significantly outperforms other models with an F1 score of 0.87, demonstrating its superior capability in capturing complex dependencies in the data. 
Random Forest follows with a respectable F1 score of 0.67, making it this study's best-performing traditional machine-learning model. 
MobileNet and its variants show moderate performance. MobileNet achieves an F1 score of 0.63, while MobileNet V2 and V3 lag behind. 
The Decision Tree and Support Vector Machine models perform similarly, with F1 scores of 0.54 and 0.53, respectively. VGG16 records the lowest F1 score at 0.31, indicating its limited effectiveness for this task.

\begin{figure}[t!]
  \centering
  \includegraphics[width=1.0\textwidth]{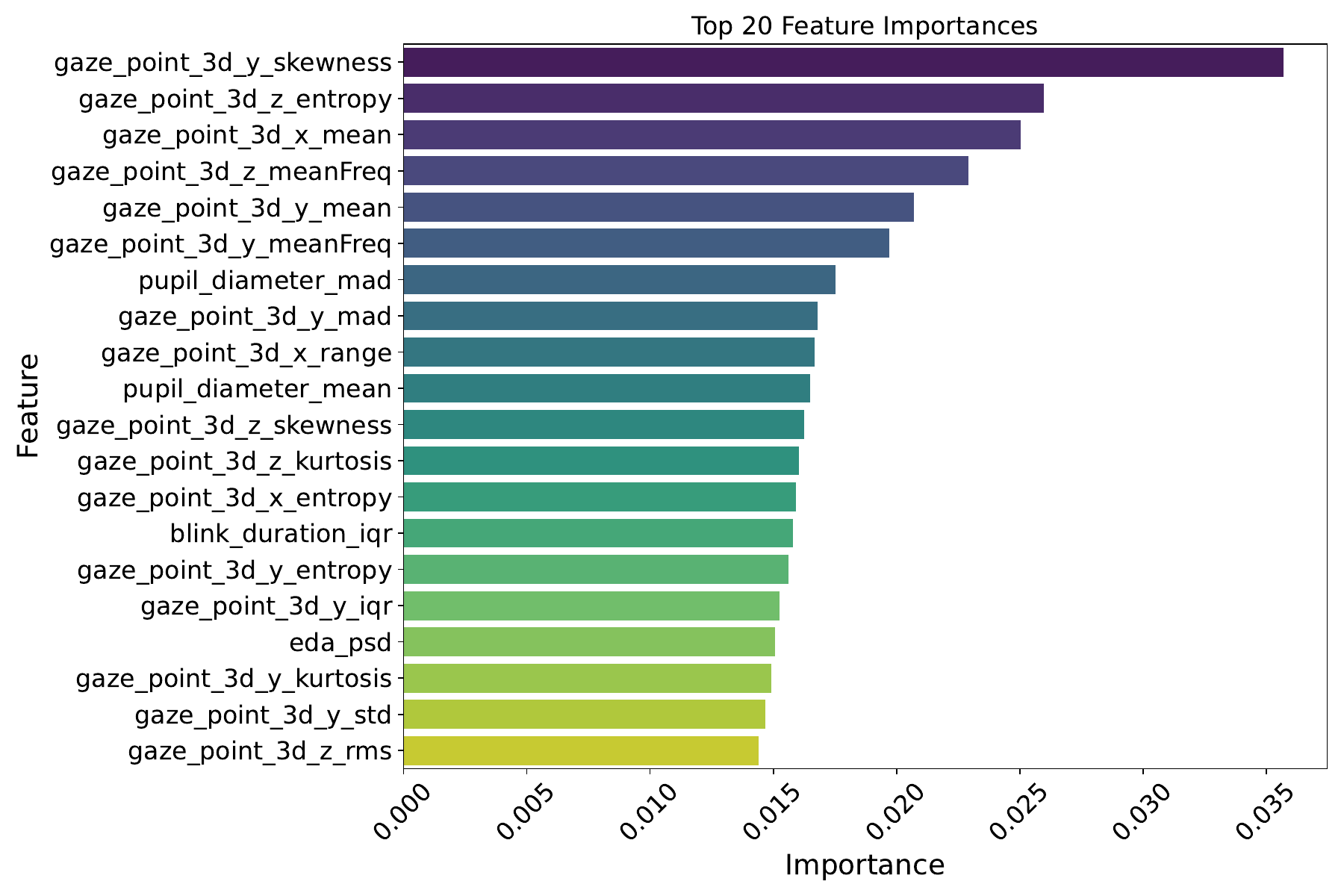}
  \caption{Feature importance when applying Random Forest}
  \label{fig:fi_rf}
\end{figure}

Confusion matrices of the best-performing models are shown in Figure~\ref{fig:confusion_matrices}.
The Transformer model shows a higher F1 Score in classifying instances correctly, as evidenced by its confusion matrix.
As the results show, comfortable distance is estimated to be better than discomfort distance.

Figure~\ref{fig:fi_rf} shows the feature importance when applying Random Forest.
The Gini impurity calculates the feature importance.
As the result shows, the eye-tracking data is more important than the wristband sensor data.
One reason is that eye-tracking data is more sensitive to distance variation than wristband sensor data.
For example, when the distance is closer to the participant, the eye-tracking data shows a smaller pupil diameter~\cite{sulutvedt2018gaze}.
In this study, participants labeled uncomfortable distance as 0.5m, the closest distance.
Therefore, the model predicts uncomfortableness regarding the distance variation.
Also, the gaze point was an important feature of the model.

Wristband sensor data was not significant in this study.
Among all features, the EDA was the most important feature of the model.
This can be because the EDA is more sensitive to physiological change than other features.
EDA is a measure of the skin's electrical activity, which is closely related to the autonomic nervous system~\cite{desai2021electrodermal}.
The autonomic nervous system controls the body's response to stress and other stimuli, and EDA is a reliable indicator of this response.

In summary, the eye-tracking data is more important than the wristband sensor data for estimating comfortable personal space.
We achieved a F1 score of 0.87 with the Transformer model, which is the best-performing model in this study.

\section{Limitations and Future Work}
The number of participants in the study is limited, which may affect the generalizability of the results. 
In this study, we only used ten participants.
The number of participants is not enough to generalize the results to the general population.
Additionally, the study does not account for individual differences such as personality traits, cultural background, gender, and situational context, which could influence the outcomes.
This may allign to the bias of the participants.

The trial size in this study is also relatively small, as we only collected two minutes of data per trial for each participant. 
This limited duration may not capture the full range of variability in the participants' responses and could affect the robustness and reliability of the findings.
The data size can be larger by asking participants to perform the task for longer periods.

To improve the robustness and reliability of the findings, future studies should consider applying a sliding window approach with longer durations and varying overlap intervals. 
This method would allow for a more comprehensive data analysis by capturing a wider range of variability in the participants' responses over extended periods.
The larger data size can achieve in applying different size of window for the sliding window approach.

The task conducted in this study does not reflect natural conditions, as participants were fully aware that they were part of an experiment, which may have influenced their behavior and responses. 
Ideally, such experiments should be conducted in more naturalistic settings, often called "in the wild," to obtain more genuine and ecologically valid data. 
Additionally, the experiment did not consider the different types of discussions that participants might engage in, which could have varying impacts on their physiological and psychological responses.

The sensors utilized in this study are restricted to two types: eye-tracking glasses, which monitor and record the participants' gaze and pupil diameter, and a wristband sensor, which measures various physiological signals such as electrodermal activity (EDA) and heart rate. 
This limitation in sensor variety may affect the comprehensiveness of the data collected, as other potentially relevant physiological and behavioral signals are not captured.

To enhance the generalizability of the results, it is essential to increase the sample size and include a more diverse participant pool in future work. 
Incorporating a broader range of individual differences, such as personality traits and cultural backgrounds, will provide a more nuanced understanding of the factors influencing outcomes. 
Extending the data collection duration and employing advanced data analysis techniques, like the sliding window approach, will help capture a more comprehensive range of participant responses. 
Conducting experiments in naturalistic settings will improve ecological validity, and expanding the variety of sensors used will ensure a more holistic capture of physiological and behavioral signals. 
These steps will contribute to more robust, reliable, and generalizable findings.

\section{Conclusion}
This study explored the estimation of comfortable personal space using multimodal sensors, integrating eye-tracking and wristband-based physiological data. 
Our findings indicate that deep learning models, particularly the Transformer model, effectively predict personal space preferences, achieving an F1 score of 0.87. Eye-tracking data plays a more significant role than wristband sensor data. 
These results highlight the potential for intelligent systems to personalize spatial arrangements in workplaces, educational institutions, and public settings, enhancing user comfort and reducing social stress. 
However, the study has limitations, including a small participant pool and controlled experimental conditions that may not fully reflect real-world scenarios. 
Future research should expand participant diversity, explore naturalistic settings, and integrate additional physiological and behavioral markers to improve model robustness. 
By advancing sensor-based estimations of personal space, this research contributes to the development of adaptive HCI that dynamically respond to individual comfort needs, paving the way for more socially aware intelligent systems.

\bibliographystyle{splncs04}
\bibliography{main}

\end{document}